\documentstyle[12pt]{article}

\advance\textheight by -5cm
\advance\textheight by -2.5cm 
\begin{document}
\begin{titlepage}
\begin{center}
March, 1993     \hfill    LBL-35337\\
 \hfill UCB-PTH-94/05\\
\vskip .4in
{\large \bf  Signals of Disoriented Chiral
Condensate}\footnote{This work was
supported
by the Director, Office of Energy
Research, Office of High Energy and Nuclear Physics,
Divisions of High
Energy Physics and Nuclear Physics
 of the U.S. Department of Energy under Contract
DE-AC03-76SF00098, and in part by the
U.S.\  National Science Foundation under grant PHY90-21139,
 and by the Natural Sciences and
Engineering Research Council of Canada.}
\vskip .5in
{\bf Zheng Huang}\footnote{Electronic Address: huang@theorm.lbl.gov}
 and {\bf Mahiko Suzuki}\footnote{Electronic Address:
suzuki@theorm.lbl.gov}\\
\vskip 0.05in
{\em Theoretical Physics Group,
Lawrence Berkeley Laboratory\\
 University of California, Berkeley, California 94720, USA}\\
\vskip 0.05in
{\bf Xin-Nian Wang}\footnote{Electronic Address: xnwang@nsdssd.lbl.gov}\\
\vskip 0.1in
{\em Nuclear Science Division,
Lawrence Berkeley Laboratory\\
 University of California, Berkeley, California 94720, USA}\\
\end{center}

\vskip .3in
\begin{abstract}
If a disoriented chiral condensate is created over an extended
space-time region following a rapid cooling in hadronic or nuclear
collisions,
the misalignment of the condensate with the electroweak symmetry
breaking can generate observable effects in the processes which
involve both strong and electromagnetic interactions.  We point out
the relevance of the dilepton decay of light vector mesons as a signal
for formation of the disoriented condensate.  We predict that the decay
$\rho^0\rightarrow \ell^+\ell^-$  will be suppressed and/or
the $\rho$ resonance peak widens,
while the decay $\omega\rightarrow \ell^+\ell^-$ will not be affected by
the condensate.
\end{abstract}
\end{titlepage}
\newpage
\renewcommand{\thepage}{\arabic{page}}
\setcounter{page}{2}

\section{Introduction}
Recently a proposal to observe the disoriented chiral condensate
(DCC) in high energy collisions \cite{an,bl,bj,alaska} has received
much attention.  Both numerical and analytical solutions that may
describe the formation of the DCC have been studied \cite{rw,ggp,hw}.
The basic idea of the DCC is that the collision debris will expand
outward from the center of a collision at the speed of light,
leaving behind a relatively cold interior whose vacuum has a chiral
orientation different from that of the normal vacuum in the exterior.
Unlike the quark-gluon plasma (QGP) which is usually assumed to be
in thermal equilibrium, the DCC is not expected to be in a local
thermal equilibrium. Indeed, the opposite extreme of free streaming
of the collision debris is expected to produce a colder interior
which leads to the DCC formation \cite{alaska}. Rajagopal and Wilczek
\cite{rw} suggested a quenching mechanism for formation of the DCC in
which the temperature drops suddenly due to expansion. It thus seems
that a zero temperature dynamics may be more appropriate to describe
hadron dynamics in the DCC.

Perhaps the most direct test for an ideal DCC
will be the observation of
an anomalous isospin charge distribution \cite{an,bl,bj,alaska}
\begin{eqnarray}
P(r)=\frac{1}{2\sqrt{r}} \label{1}
\end{eqnarray}
where $r=n_{\pi^0}/(n_{\pi^0}+n_{\pi^\pm})$.
This prediction results from a long-range correlation in the isospin
direction for a given event.  It has been demonstrated
numerically \cite{rw,hw} that a
quench after collisions may actually yield a distinct cluster structure
in the rapidity distribution which indicates a strong correlation in
the isospin direction. In order for the DCC to be interesting, its size
must be much larger than an inverse pion mass.
But we do not know at present how long the correlation length will be
for a typical DCC domain formed in hadronic or nuclear collisions.
If the correlated domain is not large enough as compared to the
total interaction region, especially in the central region of
heavy ion collisions, the expected cluster structure in momentum
phase space could be smeared out by the fluctuations of many domains.
The distribution of the total neutral to charged pion ratio
would have a binomial behavior rather than that of Eq.~(\ref{1}).
It is thus important to study
the consequences of DCC with both large and small correlated domains.

In this paper we search for a possible signature of formation of the
DCC, assuming that its space-time spread grows much larger than $1/m_\pi$.
We will focus on the misalignment of the DCC with the direction of
the electroweak symmetry breaking.  When the hadronic resonances decay
inside the DCC, their electromagnetic decay modes show a clear signal
of the disorientation of the background.  In contrast, purely hadronic
decay modes are not affected by the disorientation in the limit of a
small explicit chiral symmetry breaking and a slow space-time variation
of the DCC. We shall study the dilepton decays of the
$\rho$ and the $\omega$, since the DCC affects the $\rho$ and $\omega$
 mesons quite differently because of their different chiral properties.
In the limit of a slowly varying DCC,
the $\rho$ resonance peak in the dilepton mass spectrum
will be reduced by half. When the DCC varies rapidly in space-time,
the $\rho$ resonance peak will
be smeared and may even disappear. On the other hand,
the $\omega$ resonance will be little affected by the DCC.
Therefore there is a chance to test the formation of the DCC by
carefully measuring the dilepton decays of the $\rho$ and the $\omega$
meson. We shall first study the limit of a slowly varying DCC,
then include the space-time variation of the condensation.

\section{Vacuum Misalignment}
The DCC is created in a certain spatial domain, expands
outward and eventually disappears.  The vacuum orientation of the DCC
labeled by an isovector rotation angle $\mbox{\boldmath $\theta$}$ is
in general space-time dependent, satisfying a boundary condition:
 $\mbox{\boldmath $\theta$} = 0$ for $|{\bf x}| > t$ (space-like)
and $\mbox{\boldmath $\theta$}\rightarrow 0$  for $t\rightarrow \infty$.
Since such a configuration always has a higher energy than the true
vacuum, the DCC vacuum decays into the true vacuum semiclassically
by a coherent emission of pions, leading to the characteristic
isospin charge distribution of (\ref{1}).  It is our hope that the
DCC will live much longer than the typical time scale of low-energy
strong interactions, and the DCC spreads over a spatial region much
larger than $1/m_\pi$. Given a lifetime and a spatial size,
the condensate may also vary with space-time, which determines the
correlation  of the emitted pions.
An idealized limit often discussed in literature is an infinitely
large DCC which corresponds to an approximately homogeneous
vacuum in the cool interior.  We shall first study
the DCC vacuum in this limit
where the orientation angle $\mbox{\boldmath $\theta$}$ is
space-time independent. We later extend our argument to the DCC of a
finite size and also with a space-time variation in the interior.

Let us consider the DCC in the case of two flavors ($u$ and $d$)
in which the quark condensate matrices take the form
\begin{eqnarray}
\langle \bar{q}_L q_R\rangle =-v^3\exp
[{i 2\mbox{\boldmath $\theta$}\cdot \mbox{\boldmath $\tau$}}]\quad
;\quad  \langle \bar{q}_R q_L \rangle =-v^3\exp
[{-i2\mbox{\boldmath $\theta$}\cdot \mbox{\boldmath $\tau$}}]\; .
\label{2}
\end{eqnarray}
The DCC vacuum is related to the normal one through a chiral
rotation $U_L$ and $U_R$ with
$U_L^{\dagger}U_R=\exp [{i2\mbox{\boldmath $\theta$}
\cdot\mbox{\boldmath $\tau$}}]$. It transforms the quark fields as
\begin{eqnarray}
q_L\rightarrow U_Lq_L\quad ;\quad q_R\rightarrow U_Rq_R \; .\label{3}
\end{eqnarray}
If the quark mass is ignored, the QCD vacua would be infinitely
degenerate with respect to the chiral rotation. In this limit all
strong interactions would take exactly the same form whichever
vacuum is realized.  When the quark mass is included, the degeneracy is
lifted and the DCC vacuum
energy density is
\begin{eqnarray}
E^{\rm QCD}(\mbox{\boldmath $\theta$})
& = & \langle \mbox{Tr} (\bar{q}_LU^\dagger _LMU_Rq_R)\rangle
+\mbox{c.c.}\nonumber\\
& = & -m^2_\pi F_\pi^2\cos 2\theta\; ,
\label{4}
\end{eqnarray}
where $\theta =|\mbox{\boldmath $\theta$}|$, and $M$ is the
diagonal mass matrix with eigenvalues $m_u$ and $m_d$.
$E^{\rm QCD}(\mbox{\boldmath $\theta$})$ is minimized at
$\mbox{\boldmath $\theta$}=0$, that is,
in the $\sigma$ direction.
However, this energy shift does not play an important role in
the qualitative picture of the DCC.  The strong interaction
processes alone cannot tell of the chiral orientation of the
condensate in the limit of the small current quark mass.

On the other hand, the misalignment of the DCC vacuum with the
$\sigma$ direction will be clearly seen through the electroweak
interactions of hadrons.  In the DCC where the quark fields are
rotated as in (\ref{3}), the electromagnetic current of the $u$
 and $d$ quark is represented in the form
\begin{eqnarray}
J^{\rm EM}_\mu \rightarrow \hat{J}^{\rm EM}_\mu
=\bar{q}_L\gamma_\mu U_L^{\dagger}QU_L q_L +
\bar{q}_R\gamma_\mu U_R^\dagger QU_R q_R \; ,\label{5}
\end{eqnarray}
where $Q$ is the diagonal charge matrix with eigenvalues
$2/3$ and $-1/3$, and the quark fields are in the basis where
$\langle \bar{q}_L q_R\rangle_{\rm DCC} =
\langle \bar{q}_R q_L\rangle_{\rm DCC} =-v^3$, i.e., in a basis in
which the standard hadron mass spectroscopy is valid.
The disorientation also raises the vacuum energy of the
EM interaction, which is obtained by taking an expectation
value of the effective Hamiltonian \cite{8}
\begin{eqnarray}
\langle{\cal H}\rangle_{\rm DCC}
=-\frac{e^2}{2}\int d^4x D^{\mu\nu}(x) \langle\mbox{T} \hat{J}^{\rm EM}_\mu (x)
\hat{J}^{\rm EM}_\nu (0)\rangle\; ,
\end{eqnarray}
where $D^{\mu\nu}$ is the photon propagator.
When the explicit chiral symmetry breaking is ignored,
the electromagnetic vacuum energy density is calculated
\begin{eqnarray}
E^{\rm EM}(\mbox{\boldmath $\theta$})-E^{\rm EM}(0)=
\frac{F^2_\pi}{2}(n^2_1+n^2_2)\sin ^22\theta
 (m^2_{\pi^+}
-m^2_{\pi^0})\; , \label{6}
\end{eqnarray}
where $n_i$ ($i=1,2,3$) are the components of the unit vector
$\mbox{\boldmath $n$}=\mbox{\boldmath $\theta$}/\theta$.
The true vacuum at $\mbox{\boldmath $\theta$} = 0$ simultaneously
minimizes both $E^{\rm QCD}(\mbox{\boldmath $\theta$})$ and
$E^{\rm EM}(\mbox{\boldmath $\theta$})$.  Therefore the normal
QCD vacuum aligns with the direction of the spontaneous symmetry
breaking of the electroweak interactions, which is fixed by the
vacuum expectation value of the Higgs field.  However, the DCC
vacuum does not align with the electroweak vacuum.  One immediate
consequence of this misalignment is that the vector-meson
dominance no longer holds for the EM current in the familiar way.
In the normal vacuum, the isovector part of the EM current is
dominated by the $\rho^0$ meson ($770$ MeV).  In the DCC, the
isovector current will be dominated by the mesons that would
dominate the vector and axial-vector isospin
currents in the normal vacuum.  They are the
 $\rho$ meson and the $a_1$ meson ($1260$ MeV).
The isovector EM current in DCC
will show two resonance peaks, one at $770$ MeV and the other at
$1260$ MeV.  On the other hand, the isoscalar
current is dominated by the $\omega$ meson ($780$ MeV) and the
$\phi$ meson ($1020$ MeV) in the normal vacuum, which are
the two singlets $({\bf 1}, {\bf 1})$ of SU(2)$_L \times$SU(2)$_R$.
Therefore, no matter how much the DCC vacuum is chirally disoriented,
the $\omega$ and $\phi$ mesons are not affected at all.
This difference between the isovector and isoscalar currents
will result in very different behaviors of the dilepton decays
of the $\rho$,  $\omega$ and  $\phi$ mesons.

\section{Dilepton Decays of $\rho$ and $\omega$}
In the normal vacuum, the dilepton decays of $\rho^0$ and $\omega$
occur through one photon with the effective couplings
\begin{eqnarray}
{\cal L}_{\gamma V} =  \frac{e}{2g_\rho}F^{(\gamma )}_{\mu\nu}
F^{(\rho )\mu\nu} + \frac{e}{6g_\omega}F^{(\gamma )}_{\mu\nu}
F^{(\omega )\mu\nu}\; ,\label{7}
\end{eqnarray}
where $F_{\mu\nu}$ is the field strength.  On the mass-shell of
$\rho^0$ and $\omega$, we can rewrite (\ref{7}) into the
 effective interactions
\begin{eqnarray}
{\cal L}_{\gamma V}=\frac{em^2_\rho}{g_\rho}\rho^0_\mu A^\mu
+\frac{em^2_\omega}{3g_\omega} \omega_\mu A^\mu\; .\label{8}
\end{eqnarray}
In the DCC, the meson state with which the isovector current
couples is no longer a single meson state of mass $770$ MeV but
a linear combination of the states with mass $770$ MeV and $1260$ MeV
($= m_{a_1}$).  When the $\rho$ and $\mbox{a}_1$
mesons dominate the vector and
axial-vector isospin currents ${\bf J}_\mu$ and
${\bf J}_{5\mu}$, respectively, it holds in
the normal vacuum \cite{9} that
\begin{eqnarray}
\langle 0|{\bf J}_{5\mu}|{\bf a}_1\rangle =
\langle 0|{\bf J}_\mu |\mbox{\boldmath $\rho$}\rangle
=\frac{m^2_\rho}{g_\rho}\epsilon_\mu\; .\label{9}
\end{eqnarray}
Then the effective photon-meson couplings through the isovector
current in the DCC are given by
\begin{eqnarray}
{\cal L}^{\rm DCC}_{\gamma V} & = &
\frac{em^2_\rho}{g_\rho}{\rm Tr}\left\{
\frac{\tau_3}{2}\left[U^\dagger _L(\mbox{\boldmath $\rho$}_\mu
-{\bf a}_\mu )\cdot \mbox{\boldmath $\tau$} U_L+
U^\dagger _R(\mbox{\boldmath $\rho$}_\mu
+{\bf a}_\mu )\cdot \mbox{\boldmath $\tau$} U_R\right]\right\}A^\mu
+\frac{em^2_\omega}{3g_\omega} \omega_\mu A^\mu \nonumber\\
& =& \frac{em^2_\rho}{g_\rho}\left[\sum_{i=1,2,3}C_\rho^i(
 \mbox{\boldmath $\theta$}_L, \mbox{\boldmath $\theta$}_R)\rho^i_\mu +
\sum_{i=1,2,3}C_a^i(
 \mbox{\boldmath $\theta$}_L, \mbox{\boldmath $\theta$}_R)a^i_\mu
\right]A^\mu
+\frac{em^2_\omega}{3g_\omega} \omega_\mu A^\mu
\; .\label{10}
\end{eqnarray}
The $\mbox{\boldmath $\rho$}_\mu$ and the ${\bf a}_\mu$ in (\ref{10})
are the mass eigenstates of $\sim 770$ MeV and $\sim 1260$ MeV
 in the DCC, respectively, which are parity mixture states if a
``parity'' is defined in the normal vacuum.  The same result can be
obtained by assuming explicitly that the $\rho$ and
 $a_1$ fields transform as $\mbox{\boldmath $\rho$}\pm {\bf a}
\sim ({\bf 3},{\bf 1}) \pm ({\bf 1},{\bf 3})$ under the chiral rotation.
To derive Eq.~(\ref{10}), one only needs a chiral transformation
property of the isovector EM current, and the consequence (\ref{10})
of a vector- and
axial-vector-meson
dominance for the chiral SU(2) currents.  The latter
requires that the $\rho$ and $a_1$ fields have the same transformation
properties as the currents, i.e., as $({\bf 3}, {\bf 1}) \pm
({\bf 1}, {\bf 3})$, leading effectively to a field-current
identity \cite{10}.

All possible charge states of the $\rho$ and $a_1$ mesons
couple with the photon in the DCC with a general orientation.
A violation of the charge conservation in these transitions is
superficial.  Since the DCC vacuum may be charged, an apparent
violation of charge conservation in a photon-meson transition
is compensated by a charged decay of the vacuum that follows.
It is reasonable to expect that when the DCC is created, all
possible orientations of the chiral SU(2)$_L \times$SU(2)$_R$
are allowed with an equal probability.  We average the photon-meson
transitions in the group-parameter space of SU(2)$_L \times$SU(2)$_R$.
 On average, all six transitions, $\rho\rightarrow \gamma$
and $a_1 \rightarrow \gamma$, occur with an equal probability;
\begin{eqnarray}
\langle |C_\rho^i(
 \mbox{\boldmath $\theta$}_L, \mbox{\boldmath $\theta$}_R)|^2\rangle
=\langle |C_a^j(
 \mbox{\boldmath $\theta$}_L, \mbox{\boldmath $\theta$}_R)
 |^2\rangle=\frac{1}{6}\; . \quad\quad (i,j=1,2,3) \label{11}
\end{eqnarray}
This means that the $\rho$
peak in the $\ell^+\ell^-$ mass plot will be reduced by half
if all three charge states of $\rho$ are produced with an equal
rate, and a
new broad bump of width $\simeq 300$ MeV ($= \Gamma_{a_1}$)
may appear at 1260 MeV
\begin{eqnarray}
\Gamma\left( \rho(770 \mbox{MeV}) \rightarrow \ell^+\ell^-\right)_{\rm DCC}
& = &
\frac{1}{2} \Gamma\left( \rho^0 \rightarrow
\ell^+\ell^-\right)_{\rm normal}\; ,\\
\Gamma\left( a_1(1260 \mbox{MeV}) \rightarrow
\ell^+\ell^-\right)_{\rm DCC}
& = & \left( \frac{m_\rho}{m_{a_1}}\right) ^3
\Gamma\left( \rho(770 \mbox{MeV}) \rightarrow \ell^+\ell^-\right)_{\rm DCC}
\label{12}
\end{eqnarray}
This prediction is little affected by the current quark mass.
It is a consequence of the misalignment of the DCC with the
electroweak vacuum fixed by the Higgs field, so it would disappear
only at $F_\pi \rightarrow 0$ in which a distinction between the
$\rho$ and the $a_1$
ceases to be meaningful.

In a real experimental environment, the reduction of the $\rho$
peak will be lessened by the probability of the DCC formation in
collisions.  Note, however, that the averaging in Eq.~(\ref{11}) is done
over all possible states including the normal vacuum as a special
case and those vacua that are only slightly different from it.
If we can select the likely candidates for the DCC events by the isospin
charge asymmetry \cite{an,bl,bj,alaska}, the reduction of the
$\rho$ peak should be even larger.  However,
if the $\rho$ and $\omega$ mesons are
copiously produced together with
the coherent pions emission at the time of the decay of the
DCC itself, the test may become difficult.

The above analysis has so far been based on the idealized limit of
an infinitely large and slowly varying DCC.
Though very little is known about the detailed properties of the DCC,
 we should examine how the argument is modified by the space-time
variation of the DCC.

\section{Space-time Variation Effects}
When the size of the DCC is finite in space-time, it can happen
that the vacuum acts like a reservoir of energy and momentum for hadrons
and leads to an apparent violation of energy and momentum
conservation in their decay and
scattering processes.  The energy
and momentum transfer may become larger
when  the DCC also has a space-time variation in the interior.
For a ``realistic'' DCC, we must take into account  these space-time
variations of the DCC.  The magnitude of the transfer, when it occurs,
is determined by the uncertainty principle:
\begin{eqnarray}
\Delta p \sim 1/\Delta x \quad\quad   ;    \quad\quad
\Delta E \sim  1/\Delta t \; ,
\label{13}
\end{eqnarray}
where $\Delta x$ and  $\Delta t$ are the space and time variations.
When the DCC of a finite size is slowly varying in the interior,
 $\Delta x$ and  $\Delta t$ will be the space-time size of the DCC.  For
the DCC with a rapid space-time variation in the interior,  $\Delta x$
 and  $\Delta t$ are determined by the scale of the variation.
A DCC is meaningful and interesting only when its size is much larger
than $1/m_\pi$, while the space-time variation in the interior may be
as large as $1/m_\pi$.
Let us parametrize  $\Delta x$ and  $\Delta t$ as
\begin{eqnarray}
 \Delta x \simeq \Delta t \equiv \frac{\kappa}{m_\pi}\; .\label{14}
\end{eqnarray}
This amounts to a violation of energy-momentum conservation with
a magnitude  $\Delta p\simeq \Delta E \leq m_\pi/\kappa$ in
decay processes and in final-state interactions, if they exchange
energy-momentum with the vacuum.
The value of $\kappa$ ranges from
O(1) to $\sim 5$ or more, depending on whether it is determined by
the variation in the interior or by the size of the DCC.  We will
examine how this space-time dependence changes the previous
oversimplified picture.

     To be concrete, we describe $\rho$, $a_1$, and
$\omega$ as the massive gauge-like bosons of SU(2)$_L \times$SU(2)$_R$
 and of U(1)$_V$.  It has been long recognized that this is
phenomenologically a good description of the spin-one mesons \cite{10}.
  The relevant Lagrangian is that of a gauged $\sigma$ model to
which the vector meson masses are added:
\begin{eqnarray}
{\cal L} & =&\mbox{Tr}\mbox{D}_\mu\Phi ^\dagger \mbox{D}_\mu\Phi
-\frac{1}{4}\mbox{Tr}(F_{L\mu\nu}F_L^{\mu\nu} +F_{R\mu\nu}F_{R}^{\mu\nu})
-\frac{1}{4} \omega _{\mu\nu}
\omega _{\mu\nu}\nonumber \\
 & & +\frac{m^2_\rho}{2}\mbox{Tr}(V_{L\mu}V_L^\mu + V_{R\mu}V_R^\mu )
+\frac{m^2_\omega}{2} \omega _\mu\omega^\mu -U_\omega
-U(\Phi ^\dagger\Phi )
\label{15}
\end{eqnarray}
where $\Phi =\frac{1}{2}(\sigma +i\mbox{\boldmath $\pi$}\cdot
\mbox{\boldmath $\tau$})$,
$V_{L(R)\mu}=\frac{1}{2}(\mbox{\boldmath $\rho$}_\mu\pm
\mbox{\boldmath $a$}_\mu)\cdot
\mbox{\boldmath $\tau$}$ and
\begin{eqnarray}
\mbox{D}_\mu\Phi =\partial_\mu\Phi +igV_{L\mu}\Phi
-ig\Phi V_{R\mu}\; .\label{16}
\end{eqnarray}
In (\ref{15}), $U_\omega$ is the anomaly term that generates the
$\omega\rightarrow \pi\pi\pi$ decay amplitude, and
$U(\Phi^\dagger\Phi )$ denotes the mass and potential terms of the $\Phi$.
Upon spontaneous symmetry breaking, the $a_1$ meson acquires an
additional mass. The explicit symmetry breaking term
$-m^2_\pi F_\pi\sigma$
is left out for the moment. We will discuss its effect later.

For the DCC with the space-time dependent angle parameters
$\mbox{\boldmath $\theta$}_L(x)$ and $\mbox{\boldmath $\theta$}_R(x)$,
the quarks fields $q'(x)$ in the DCC are related to the quark fields
$q(x)$ in the normal vacuum through
\begin{eqnarray}
 q_L  =  U_L(x)q'_L  \quad\quad ;\quad\quad    q_R = U_R(x)q'_R,
\label{17}
\end{eqnarray}
where $U_L(x)$ and $U_R(x)$ depend on space and time instead of
global transformations in Eq.~(\ref{3}).  The
same transformation relates the meson field
$\Phi '$ in the DCC to the meson field $\Phi$ in the normal vacuum as
\begin{eqnarray}
\Phi = U_L(x)\Phi 'U^\dagger _R(x)\quad ;\quad
\Phi^\dagger =  U_R(x)\Phi '^\dagger U^\dagger_L (x)\; , \label{18}
\end{eqnarray}
where $\langle \Phi\rangle=F_\pi /2$ in the normal vacuum.  In a
space-time dependent DCC, the spin-one fields also develop vacuum
expectation values since the background carries angular momentum.
One convenient basis for the spin-one fields is obtained by the
local transformations,
\begin{eqnarray}
V_{L\mu}& = & U_L(x) V '_{L\mu}U^\dagger _L(x)
-\frac{i}{g}U_L(x)\partial_\mu U^\dagger _L(x)\; \nonumber\\
V_{R\mu}& = & U_R(x) V '_{R\mu} U^\dagger _R(x)
-\frac{i}{g}U_R(x)\partial_\mu U^\dagger _R(x)\; ,\label{19}\\
\omega_\mu & = & \omega '_\mu\nonumber
\end{eqnarray}
where $\langle V_{L(R)\mu}\rangle =0$ in the normal vacuum.
Since (\ref{19}) is an SU(2)$_L\times$SU(2)$_R$ gauge transformation,
the Lagrangian in(\ref{15}) is invariant except for the $\rho$-$a_1$
mass term
\begin{eqnarray}
{\cal L}& \rightarrow & {\cal L}+\delta{\cal L}\;,\nonumber\\
\delta {\cal L}& = &  i\frac{m^2_\rho}{g}\mbox{Tr}
\left[ U^\dagger _L(x)\partial^\mu
U_L(x)V '_{L\mu}\right]
+ i\frac{m^2_\rho}{g}\mbox{Tr}\left[ U^\dagger _R(x)\partial^\mu
U_R(x)V '_{R\mu}\right]\label{20}\\
& & +\frac{m^2_\rho}{2g^2}\mbox{Tr}\left[ \partial^\mu U_L(x)\partial_\mu
U^\dagger_L(x)\right]
+\frac{m^2_\rho}{2g^2}\mbox{Tr}\left[ \partial^\mu U_R(x)\partial_\mu
U^\dagger_R(x)\right]\; .\nonumber
\end{eqnarray}
In this basis, $\delta {\cal L}$ contains a space-time dependent
tadpole interaction for $\rho$ and $a_1$.  The presence of these
tadpoles means that $\rho$ and $a_1$ acquire effectively
space-time dependent mass in the DCC.   One way to understand this
interpretation
 is to sum  an infinite series of the tadpole diagrams in the
$\rho$ and $a_1$ propagators.  Alternatively we can eliminate the
tadpole terms by using the exact classical equations of motion for the
$\rho$ and $a_1$ fields in the presence of the $\Phi$ field.
  This amounts to an additional shift $\langle V '_{L(R)\mu}\rangle$
 for the $\rho$ and $a_1$ fields in addition to (\ref{19}), which
generates additional vector-meson mass terms through the interaction
$g^2V_\mu V_\nu V^\mu V^\nu$, contained in
$\mbox{Tr}(F_{\mu\nu}F^{\mu\nu})$. We can see from $\delta {\cal L}$
 that the DCC expectation value of $V '_{L(R)\mu}$
is related to the space-time variation of the DCC
\begin{eqnarray}
\langle V '_{L(R)\mu}\rangle_{\rm DCC} =
\mbox{O}\left(\langle U^\dagger _{L(R)}(x)\partial_\mu
U_{L(R)}(x)\rangle_{\rm DCC}\right)\simeq \frac{m_\pi}{\kappa}\; .
\label{21}
\end{eqnarray}
This additional shift generates also a new
$\rho\rightarrow \pi\pi$ interaction with the coupling of
O($g^2m_\pi/\kappa m_\rho$) through
$g^2(\langle \mbox{\boldmath $\rho$}_\mu\rangle\times
\mbox{\boldmath $\pi$})\cdot ( \mbox{\boldmath $\rho$}^\mu\times
\mbox{\boldmath $\pi$})$.  This is however very small when $\kappa \geq 1$.
  The space-time dependent part of the
mass is $\Delta m_V^2\sim g^2\langle V_\mu
\rangle\langle V^\mu\rangle$.  With $g^2/4\pi = 2.5$ from the
$\rho\rightarrow \pi\pi$  decay
width, the space-time dependent smearing of the $\rho$ mass is of the
order of $\Delta m_\rho^2 \sim 30m_\pi^2/\kappa^2$.
The smearing in mass square translates to a smearing in mass itself as
\begin{eqnarray}
   \delta m_\rho  \leq   \frac{30m_\pi^2}{2\kappa^2m_\rho}\; .\label{22}
\end{eqnarray}
If the DCC is slowly varying and the size of the DCC determines
$\kappa$ ($\geq 5$), $\delta m_\rho\leq 25$ MeV; the $\rho$ meson
will maintain its resonance shape with a
suppressed $\ell^+\ell^-$ event number.
  When the space-time variation in the interior is rapid ($\sim \mbox{O}
(1/m_\pi)$), the $\rho$ resonance becomes very wide
due to the smearing effect,
and may virtually disappear if most of $\rho$'s are created in contact
with the vacuum.  The effective $\rho$-$\gamma$ transition
coupling also acquires a space-time dependence, which
adds to an apparent violation
of energy-momentum conservation in the decay
$\rho\rightarrow \ell^+\ell^-$.  Therefore, in the
case of a rapidly varying DCC, one
may expect a rather prominent shape change at the $\rho$ resonance in both
the $\pi\pi$ decay mode and the $\ell^+\ell^-$ decay mode.
In both cases, the $\omega$ field does not change
 and acts as if there were no DCC background.
In particular, the $\omega$ resonance peak of the dilepton will maintain
its shape and serves as a ``standard''. A careful comparison of the
$\rho$ and the $\omega$ components at the overlapped
$\rho$-$\omega$ peak will reveal
the existence of DCC.

We finally remark on the explicit chiral symmetry breaking that
has so far been ignored.  The main effect of this breaking is to smear the
effective pion mass square over a range from $-m^2_\pi$ to  $m^2_\pi$
in the DCC.  Due to its short lifetime, the $\rho$ decays mostly
inside the DCC.  The smearing of the pion mass has very little effect on
the $\rho$ decay width because the  phase space for $2\pi$  is large enough.
The effect on the  phase space for $3\pi$ in the decay $\omega\rightarrow
\pi\pi\pi$
 will be a little larger.  If the decay $\omega\rightarrow
\pi\pi\pi$ occurs inside the DCC, some widening of the $\omega$ width
may occur.   However, the $\omega$ lifetime is probably too long for
the $\omega$ to decay inside the DCC.  If the majority of the $\omega$
 decays outside the DCC, the $3\pi$ peak of the $\omega$
will remain relatively clean.

\section{Discussions}
We have shown that the misalignment of the DCC with the electroweak
direction  results in the suppression of the decay
$\rho^0\rightarrow \ell^+\ell^-$
by a factor of two on average, while no suppression is expected for
$\omega\rightarrow \ell^+\ell^-$. The marked
difference between $\rho$ and
$\omega$ arises from their different chiral properties.  This is in
sharp contrast to the QGP, which affects equally the $\rho^0=
\frac{1}{\sqrt{2}}(\bar uu-\bar dd)$ and the
 $\omega =
\frac{1}{\sqrt{2}}(\bar uu+\bar dd)$ except for the large difference in
their lifetimes which favors $\rho^0\rightarrow \ell^+\ell^-$ over
$\omega\rightarrow \ell^+\ell^-$ during the hadronic
expansion \cite{heinz}.
Some may question whether the $\rho$ and the $\omega$ can be produced
at all in the QGP.  It appears that the DCC is a less hostile
environment for these resonance to be formed.  In any case, we have to
learn about it from experiment.  It is clearly necessary to carry out
more detailed theoretical analysis of the outgoing dilepton spectrum
for the space-time dependent DCC of which  we have little
understanding at present.

In order to carry out our test in experiment, we need to know the
relative production rate of $\rho$ and $\omega$ in hadron
collisions.  At very low energies, it is known \cite{11} that the
production rates for the $\rho^0$ and the $\omega$ are equal in pp
collision: $\sigma (\rho^0 )/\sigma (\omega )= 1.0\pm 0.2$ at
$\sqrt{s} = 5$ GeV;
$\sigma (\rho^0 )/\sigma (\omega )= 1.07\pm 0.2$ at
$\sqrt{s} = 6.8$ GeV.
Recently,  the cross sections were measured at higher energy
by NA27 experiment \cite{12} through the hadronic decay modes.
The result is $\sigma (\rho^0)=(12.6\pm 0.6)$ mb
{\em vs}.\  $\sigma (\omega ) = (12.8  \pm 0.8)$ mb in pp collision at
$\sqrt{s} = 27.5$ GeV.  The quark model predicts that these cross
sections ought to be equal.  Though an independent check for their
separate cross sections would be desirable,
the equality of the production rates
appears to be a very safe assumption.

A search for the Centauro events has been made by UA5 \cite{13} and
UA1 \cite{14} at $\sqrt{s} = 546$ GeV and by UA5 \cite{15}
 at $\sqrt{s} = 900$ GeV.
An upper limit was set on the Centauro production by UA5 \cite{15}
 at the level of a few times $10^{-3}$ per inelastic events at
$\sqrt{s} = 900$ GeV.  Since an ideal DCC with large correlated domains
will generate
the Centauro and anti-Centauro events, there has been so far no evidence
for clear DCC events.
We hope that our suggestion made in this paper will be considered
in future experiments  searching for DCC,
in both hadronic and heavy nucleus collisions.

\vspace{12pt}
\begin{center}
{\bf Acknowledgements}
\end {center}

    We are grateful to J.\ D.\ Bjorken for stimulating conversations.
This work was supported in part by the Director, Office of Energy Research,
Office of High Energy and Nuclear Physics, Divisions of High Energy Physics
and Nuclear Physics of the U.S.\
Department of Energy under contract DE-AC03-76SF00098 and in part by the
U.S.\  National Science Foundation under grant PHY90-21139.
Z.H.\  acknowledges a financial support of the National Science and
Engineering Research Council of Canada.

\pagebreak


\begin{thebibliography}{99}
\bibitem{an}A.A.\ Anselm, Phys.\ Lett.\ B {\bf 217}, 169 (1989);
A.A.\ Anselm and  M.G.\ Ryskin, Phys.\ Lett.\ B {\bf 266}, 482 (1991).
\bibitem{bl}J.-P.\ Blaizot and A.\ Krazywicki, Phys.\ Rev.\ D {\bf 46},
246 (1992).
\bibitem{bj} J.\ Bjorken, Int.\ J.\ Mod.\ Phys. A {\bf 7}, 4189 (1987);
Acta Physics Polonica B {\bf 23}, 561 (1992);
J.\ Bjorken,
K.\ Kowalski and C.\ Taylor,
SLAC Preprint SLAC-PUB-6109 (1993).
\bibitem{alaska}J.\ Bjorken, K.\ Kowalski and C.\ Taylor,
SLAC Preprint, SLAC-PUB-6109 (1993).
\bibitem{rw}K.\ Rajagopal and F.\ Wilczek, Nucl.\ Phys.\ B {\bf 399},
395 (1993); B {\bf 404}, 577 (1993); F.\ Wilczek, Princeton Preprint
IASSNS-HEP-93/48 (1993).
\bibitem{ggp}S.\ Gavin, A.\ Gocksch and R.D.\ Pisarski,
Brookhaven Preprint
BNL-GGP-1.
\bibitem{hw}Z.\ Huang and X.N.\ Wang, LBL Preprint
LBL-34931 (1993).
\bibitem{8}M.E.\ Peskin, Nucl.\ Phys.\ B {\bf 185}, 197 (1981);
J.\ Preskill, Nucl.\ Phys.\ B {\bf 177}, 21 (1981).
\bibitem{9}S.\ Weinberg, Phys.\ Rev.\ Lett.\ {\bf 18}, 507 (1967)
\bibitem{10}N.\ Kroll, T.D.\ Lee and B.\ Zumino, Phys.\
Rev.\ {\bf 157},
1376 (1967);
T.D.\ Lee and B.\ Zumino, Phys.\ Rev.\ {\bf 163}, 1667 (1967).
\bibitem{heinz}U.\ Heinz and K.S.\ Lee, Nucl.\  Phys.\  A {\bf 544},
    503c (1992).
\bibitem{11}V.\ Blobel et al., Phys.\ Lett.\ B {\bf 48}, 73
(1974); Nucl.\
Phys.\ B {\bf 69}, 237 (1974).
\bibitem{12}M.\ Aguilar-Benitez et al., LEBE-EHS Collaboration,
Z.\ Phys.\
C {\bf 50}, 405 (1991)
\bibitem{13}K.\ Alpgard et al.,
UA5 Collaboration, Phys.\ Lett.\ B {\bf 115}, 71 (1982).
\bibitem{14}G.\ Arnison et al., UA1 Collaboration, Phys.\ Lett.\ B
{\bf 122}, 189 (1983).
\bibitem{15}G.J.\ Alner et al., Phys.\ Lett.\ B {\bf 180}, 415 (1986).
\end{thebibliography}
\end{document}